\documentclass[10pt,journal,compsoc]{IEEEtran}
\usepackage{amssymb}
\usepackage{subfigure}
\usepackage{algorithmic}
\usepackage[ruled,linesnumbered]{algorithm2e}
\usepackage{makecell}
\usepackage{graphicx}

\newtheorem{Def}{Definition}

\ifCLASSOPTIONcompsoc
  \usepackage[nocompress]{cite}
\else
  \usepackage{cite}
\fi
\ifCLASSINFOpdf
\else
\fi
\begin{document}
%
\title{HoSIM: Higher-order Structural Importance based Method for Multiple Local Community Detection}
%
%
%
%

\author{Boyu Li,
        Meng Wang,
        John E. Hopcroft,
        and~Kun He
\IEEEcompsocitemizethanks{\IEEEcompsocthanksitem B. Li, M. Wang and K. He are with the Department
of Computer Science and Technology, Huazhong University of Science and Technology, Wuhan, Hubei.\protect\\
\IEEEcompsocthanksitem J. E. Hopcroft is with Department of Computer Science, Cornell University, Ithaca, NY.}
\thanks{Manuscript received April 19, 2005; revised August 26, 2015.}}

%
%

\markboth{Journal of \LaTeX\ Class Files,~Vol.~14, No.~8, August~2015}%
{Shell \MakeLowercase{\textit{et al.}}: Bare Demo of IEEEtran.cls for Computer Society Journals}
%



\IEEEtitleabstractindextext{%
\begin{abstract}
Local community detection has attracted much research attention recently, and many methods have been proposed for the single local community detection that finds a community containing the given set of query nodes. However, nodes may belong to several communities in the network, and detecting all the communities for the query node set, termed as the multiple local community detection (MLCD), is more important as it could uncover more potential information. MLCD is also more challenging because when a query node belongs to multiple communities, it always locates in the complicated overlapping region and the marginal region of communities. Accordingly, detecting multiple communities for such nodes by applying seed expansion methods is insufficient.

In this work, we address the MLCD based on higher-order structural importance (HoSI). First, to effectively estimate the influence of higher-order structures, we propose a new variant of random walk called Active Random Walk to measure the HoSI score between nodes. Then, we propose two new metrics to evaluate the HoSI score of a subgraph to a node and the HoSI score of a node, respectively. Based on the proposed metrics, we present a novel algorithm called HoSIM to detect multiple local communities for a single query node. HoSIM enforces a three-stage processing, namely subgraph sampling, core member identification, and local community detection. The key idea is utilizing HoSI to find and identify the core members of communities relevant to the query node and optimize the generated communities. Extensive experiments illustrate the effectiveness of HoSIM.
\end{abstract}

}

\maketitle

\IEEEdisplaynontitleabstractindextext

%
\IEEEpeerreviewmaketitle

\IEEEraisesectionheading{\section{Introduction}\label{sec:introduction}}
\IEEEPARstart{A}{s} one of the most significant topics for network analysis, local community detection (LCD) has been widely applied in many fields successfully. Most existing LCD methods have been proposed for the single local community detection (SLCD)~\cite{KlosterG14, LuoBYLHZ20, AndersenCL06}, such that for a single or a few query nodes, we could find a community that contains all the query nodes. However, nodes may belong to multiple communities in the network, and detecting all communities for the query node set, termed as multiple local community detection (MLCD)~\cite{HeSBHL15, NiLZH20, KamuhandaH18}, is more important as it could uncover more potential information in the network. Therefore, MLCD has been attracting more attention recently.

Addressing the problem of MLCD, He et al.~\cite{HeSBHL15} divide the ego network of the query node into numerous independent connected components by removing the query node from its ego network. Then, they detect a local community for each connected component by using a local spectral method. Kamuhanda et al.~\cite{KamuhandaH18} sample a subgraph by applying breadth-first search and use the method of Non-negative Matrix Factorization (NMF) to determine the number of local communities as well as the members of each community. Hollocou et al.~\cite{HollocouBL17} iteratively pick new seeds by using scoring functions and discover the local communities based on the new seeds, but their approach requires the number of local communities as a hyper-parameter. Ni et al.~\cite{NiLZH20} construct candidates around the seeds based on local structure information and pick the representative nodes from the candidates. Then, they uncover a community for each representative node by using seed expansion method.

\begin{figure}[t]
	\centering
	\includegraphics[width=2.5in]{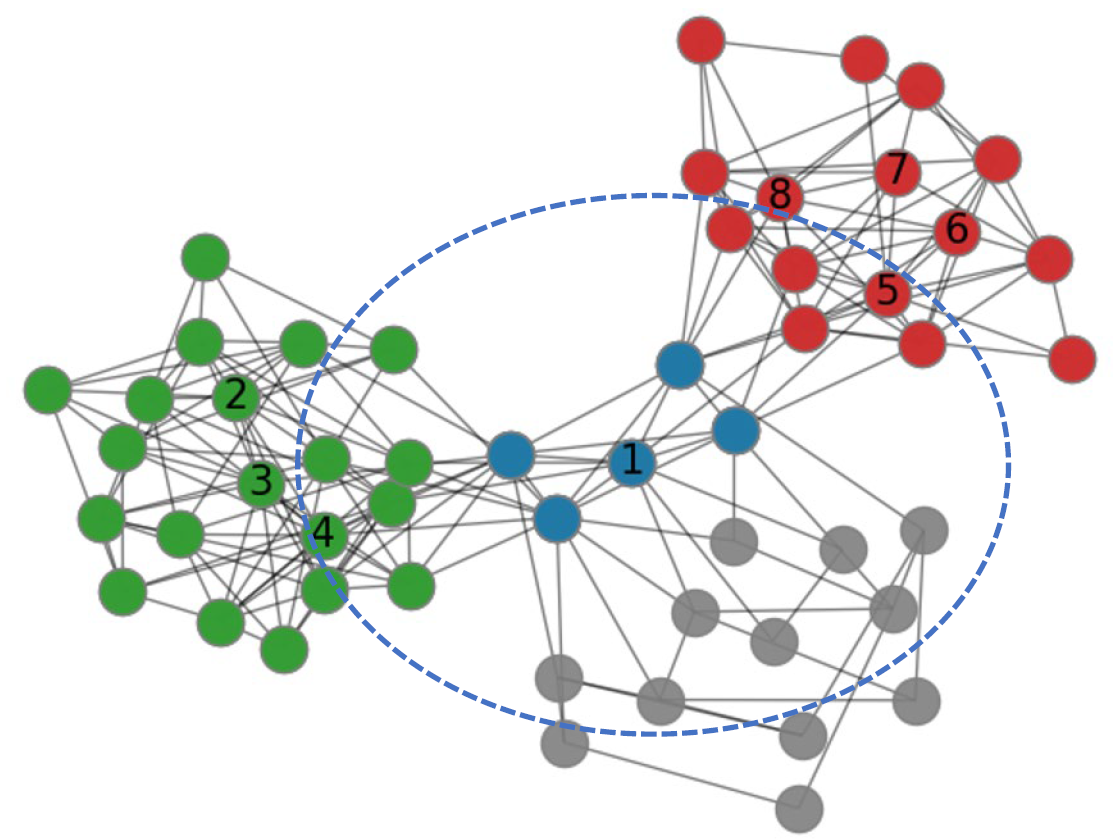}
	\caption{Multiple local community detection.}
	\label{fig_1}
\end{figure}

We observe that there are three main challenges for MLCD. The first is how to sample a subgraph that contains all nodes in the communities of the query node at the same time keeping the sampled subgraph as small as possible to speedup the followup calculation. For a query node which locates in the marginal region of the communities, finding and selecting the core members of the communities relevant to the query node as the seed nodes can effectively improve the detection results~\cite{BianHDZ20}. However, previous sampling strategies mainly focus on picking nodes related to the query node but ignore the importance of the core members. For instance, in Figure~\ref{fig_1}, the green nodes and the red nodes are the members of two different communities, the blue nodes are in the overlapping region of the two communities, while the gray nodes are close to the query node but do not belong to any community. For the query node 1 from the marginal region of the communities, the sampled subgraph with the blue dotted curve is returned by applying some well-known random walk technique. However, since the core members of the communities, i.e., nodes 2 to 4 and nodes 5 to 8, are a little far away from the query node, the core members may not be quite as related to the query node. Therefore, the core members of the communities are hard to be contained in the subgraph when the query node locates in the marginal region of the communities.

The second challenge is how to automatically and accurately determine the number of communities containing the query node. The query node must be in the complicated overlapping region when the query node belongs to multiple communities, and determining an appropriate number of communities is a precondition for high quality detection. Recent studies~\cite{BaiYS17, WangLDT16} demonstrate that the core members of a community are always connected densely and are much more distinct from other members. Thus, we can estimate the number of communities according to the sets of core members being found. For instance, in Figure~\ref{fig_1}, if we could identify the core members of the communities, i.e., nodes 2 to 4 and nodes 5 to 8, we can conclude that query node 1 belongs to two communities.

\begin{figure}[t]
	\centering
	\includegraphics[width=2.0in]{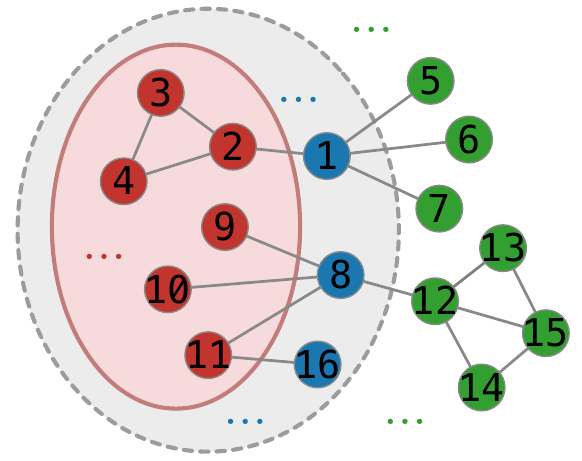}
	\caption{Importance of the higher-order structure.}
	\label{fig_2}
\end{figure}

The third challenge is that most seed expansion methods generate the local community by only optimizing the community scoring functions (e.g., conductance~\cite{ShiM00} or local modularity~\cite{LuoWP08}) but ignoring the effects of higher-order structures around the nodes. For instance, as shown in Figure \ref{fig_2}, the red nodes are members of the uncovered community at present, the blue nodes are external adjacency nodes of the community, and the green nodes are other external nodes. According to the conductance metric, node 8 is more likely to be joined into the community than node 1. However, we observe that node 1 connects a strong structure within the community (i.e., nodes 2 to 4), but node 8 links a dense structure outside the community (i.e., nodes 12 to 15). Hence, node 1 should be more possible to be a member of the community than node 8 in terms of higher-order structure.

To address the above challenges, in this work, we solve the MLCD based on the Higher-order Structural Importance (HoSI) in the graph. First, we present a new variant of random walk called Active Random Walk (ARW) to effectively measure the HoSI score between the nodes. At each step of ARW, ARW first performs a step of standard random walk on the probability vector, then we add an operation that pushes all the remaining probability of the seed node to its neighbor nodes. In this way, the neighbor nodes have more probabilities to be diffused to other nodes at the next step. Therefore, ARW is more efficient to discover the dense structures around the neighborhood of the seed node than existing random walk methods, as validated by our followup experiments. 

Then, we propose two new metrics to measure the HoSI score of a subgraph to a node and the HoSI score of a node, respectively. The first metric denotes the probabilities diffused from the node and preserved inside the subgraph. Thus, if the score of a community structure to a node is high, the node is more likely to be a member of the community structure. The second metric represents the tightness of the structure around the node. Then, a node with high score is more likely to be a core member of a community.

Based on the proposed metrics, we present a novel algorithm called \textbf{H}igher-\textbf{o}rder \textbf{S}tructural \textbf{I}mportance based \textbf{M}ethod (HoSIM) for the MLCD to address the above three challenges. The key idea is to utilize the HoSI score to find and identify the core members of communities related to the query node and optimize the generated communities. First, HoSIM samples an initial subgraph containing the most related nodes to the query node and iteratively expands the subgraph by adding the nodes with high HoSI score of the subgraph to the node. As a result, the final subgraph includes the most related nodes to the query node and the core members of communities. Then, HoSIM identifies the core members of communities inside the subgraph by picking the nodes with high HoSI score, and divides the core members into multiple disjoint sets by finding the independent connected components in the subgraph based on the core members. In this way, the number of communities is automatically determined by the number of core member sets. Finally, for each set of core members, HoSIM selects a set of high quality seeds and generates a local community by performing short random walks. Moreover, HoSIM applies addition operation and removal operation based on the HoSI scores of the community to nodes that further optimizes the generated communities.

The main advantages of HoSIM are summarized as follows: (1) the sampling strategy can effectively find the core members of local communities; (2) HoSIM can automatically determine the number of local communities according to the sets of core members being found; (3) HoSIM performs addition operation and removal operation to improve the detection quality. The extensive experiments illustrate that HoSIM greatly outperforms other baselines as evaluated by the $F_1$-score.

\section{Preliminaries}
\subsection{Problem Formulation}
Given a network modeled as an undirected and unweighted graph $G=(V,E)$, where $V$ and $E$ represent the sets of nodes and edges, respectively. Let $\textbf{A} \in \{0,1\}^{|V| \times |V|}$ be the adjacency matrix, and $\textbf{D}$ denote the diagonal matrix of the node degrees. For a probability vector $p$ and any node $u \in V$, let $p[u]$ denote the probability in $p$ for $u$. Let $\mathcal{C}$ represent the set of ground-truth communities each containing $v_q$, where $v_q \in V$ is a query node. The problem of concern is to detect a set of communities $\mathcal{C}'$ each containing $v_q$, such that for any community $C \in \mathcal{C}$, there exists a detected community $C' \in \mathcal{C}'$ satisfying $C' \equiv C$. The important notations are summarized in Table \ref{tab_not}.

\subsection{Evaluation Metric}
We utilize Jaccard $F_1$-score~\cite{He18Hicode} to evaluate the effectiveness of community detection methods. For each query node, $F_1$-score is defined as
\begin{displaymath}
	F_{1} = \frac{2 * precision_{avg} * recall_{avg}}{precision_{avg} + recall_{avg}},
\end{displaymath}
where $recall_{avg}$ indicates how well the set of ground-truth communities is detected, and $precision_{avg}$ evaluates the relevance of detected communities to ground-truth communities.

Specifically, given a query node $v_q \in V$, for a ground-truth community $C \in \mathcal{C}$, its $recall$ is calculated as
\begin{displaymath}
	recall(C) = \max_{j} \frac{|C \cap C'_j|}{|C \cup C'_j|}, C'_j \in \mathcal{C}',
\end{displaymath}
and $recall_{avg}$ is counted as
\begin{displaymath}
	recall_{avg} = \frac{\sum_{C \in \mathcal{C}}{recall(C)}}{|\mathcal{C}|}.
\end{displaymath}

In a similar way, given a query node $v_q \in V$, for a detected community $C' \in \mathcal{C}'$, its $precision$ is calculated as
\begin{displaymath}
	precision(C') = \max_{i} \frac{|C_i \cap C'|}{|C_i \cup C'|}, C_i \in \mathcal{C},
\end{displaymath}
and $precision_{avg}$ is counted as
\begin{displaymath}
	precision_{avg} = \frac{\sum_{C' \in \mathcal{C}'}{precision(C')}}{|\mathcal{C}'|}.
\end{displaymath}

\begin{table}
	\footnotesize
	\centering
	\caption{Summary of Notations.}
	\label{tab_not}
	\begin{tabular}{r|l}
		\hline\noalign{\smallskip}
		Notation & Description\\
		\hline
		$\textbf{A}$ & The adjacency matrix \\
		$\textbf{D}$ & The diagonal matrix of node degrees \\
		$HS(u, v)$ & The HoSI score of node $v$ to node $u$ \\
		$HS(u, G_{sub})$ & The HoSI score of subgraph $G_{sub}$ to node $u$ \\
		$HS(u)$ & The HoSI score of node $u$ in the graph \\
		$G_{u,l}$ & The $l$-distance ego network of $u$ \\
		$\textit{diffuse}_{l}^{(k)}(u)$ & Performing $k$ steps of random walk from $u$ within $G_{u,l}$ \\
		$p_{u,l}^{(k)}$ & The probability vector obtained by $\textit{diffuse}_{l}^{(k)}(u)$ \\
		\noalign{\smallskip}\hline
	\end{tabular}
\end{table}

\subsection{PageRank-Nibble}
In this work, we utilize the method of PageRank-Nibble~\cite{AndersenCL06} (PRN) to grow the seed nodes into a community. PRN first applies ApproximatePR to generate an $\epsilon$-approximate PageRank vector representing the probability distribution diffused from the seed node. ApproximatePR initializes two vectors $p=0$ and $r=s$, where $s$ is the starting vector. At each iteration, ApproximatePR randomly picks a node $u$ satisfying $r[u] \geq \epsilon d[u]$, where $d[u]$ is the degree of $u$, and conducts a push operation on $u$. The push operation transports $r[u]$ by three steps as follows: (1) moves an $\alpha$ fraction from $r[u]$ to $p[u]$; (2) spreads the remaining $(1-\alpha)$ fraction within $r[u]$ by applying one step of lazy random walk to the vector $(1-\alpha)r[u]\chi_u$, where $\chi_u$ is the starting vector of $u$; (3) retains the rest probability in $r[u]$. The iteration process stops if $\nexists u \in V$, $r[u] \geq \epsilon d[u]$.

After obtaining the PageRank vector, PRN sorts the nodes according to the probability-per-degree in descending order, and produces a collection of node sets each denoted as $S_{j}=\{v_1, v_2, \cdots, v_j\}$ for $j \in [1, |V|]$. The node set with the smallest conductance is selected as the community.

\section{HoSI Measurement}
\subsection{HoSI Score between Nodes}
\label{sec:hnodes}
We first define the diffusion operation in order to measure the HoSI score between nodes.

\begin{Def}[Diffusion Operation]
	Given an $l$-distance ego network $G_{u,l}$ for $u \in V$, a diffusion operation, denoted as $\textit{diffuse}^{(k)}_{l}(u)$, performs $k$ steps of random walk that diffuses the probability starting from $u$ within $G_{u,l}$ to obtain a probability vector $p^{(k)}_{u,l}$.
\end{Def}

Then, for any node $v$ inside $G_{u,l}$, $p^{(k)}_{u,l}[v]$ denotes the HoSI score of node $v$ to node $u$, termed as $HS(u,v)$.

\begin{Def}[HoSI Score between Nodes]
	For a node $u \in V$ and its $l$-distance ego network $G_{u,l}$, the HoSI score of node $v \in G_{u,l}$ to node $u$, termed as $HS(u,v)$, is measured by $p^{(k)}_{u,l}[v]$.
\end{Def}

Although numerous variants of random walks have been proposed, most of them are insufficient to measure the HoSI score between nodes. For instance, we apply Personalized PageRank (PPR) and Lazy Random Walk (LRW)~\cite{HeSBH19} to implement the short random walks for the diffusion operation, respectively, and conduct the diffusion operation on node 1 in Figure \ref{fig_2}. Table~\ref{tab_1} lists the probability distributions. For PPR with 4 steps, node 1 occupies the most probability, and few probabilities are distributed to other nodes; for LRW with 4 steps, although node 2 has much denser structure than node 5, the probability on node 2 is not obviously higher than that of node 5.

\begin{table}[ht]
	\small
	\centering
	\caption{Probabilities calculated by variants of random walk. }
	\begin{tabular}{c|c|c|c|c|c|c|c}
		\hline
		Node & 1 & 2 & 3 & 4 & 5 & 6 & 7 \\
		\hline
		PPR & 0.712 & 0.045 & 0.114 & 0.114 & 0.005 & 0.005 & 0.005 \\
		LRW & 0.419 & 0.138 & 0.049 & 0.049 & 0.115 & 0.115 & 0.115 \\
		ARW & 0.000 & 0.262 & 0.141 & 0.141 & 0.152 & 0.152 & 0.152 \\
		\hline
	\end{tabular}
	\label{tab_1}
\end{table}

In this work, we propose a new variant of random walk that can effectively measure the HoSI score between nodes with a few steps, called Active Random Walk (ARW). The definition of ARW is given as follows.

\begin{Def}[Active Random Walk] 
	Each step of active random walk (ARW) includes two stages. The first is performing a step of standard random walk on the probability vector, such that for any node $u$, we have
	\begin{displaymath}
		p^{(k)}_{u,l}=p^{(k-1)}_{u,l}\textbf{T}_{u,l},
	\end{displaymath}
	where $\textbf{T}_{u,l}=\textbf{D}_{u,l}^{-1}\textbf{A}_{u,l}$ is the corresponding transition matrix induced by $G_{u,l}$, and $p^{(0)}_{u,l}$ is the initial vector consisting of 1 for $u$ and 0 for the other nodes within $G_{u,l}$. Secondly, the probability on $u$ is pushed to its neighbor nodes, such that
	\begin{displaymath}
		p^{(k)}_{u,l}=p^{(k)}_{u,l} + p^{(k)}_{u,l}[u]\textbf{T}_{u,l}[u,:],
	\end{displaymath}
	and then
	\begin{displaymath}
		p^{(k)}_{u,l}[u]=0,
	\end{displaymath}
	where $\textbf{T}_{u,l}[u,:]$ is the vector of the row associated with $u$. 
\end{Def}

The main advantage of ARW is that the probability on the seed node is pushed to its neighbor nodes at the end of every step. Thus, the neighbor nodes have more probabilities to be diffused to other nodes at the next step, resulting that the dense structures of neighbor nodes can be effectively discovered. For instance, as shown in Table~\ref{tab_1}, for ARW with 4 steps, neighbor node 2, which has denser structure, is distinct from other neighbor nodes in terms of the probability.

In this work, we set $l=2$ because according to $\gamma$-decaying~\cite{ZhangC18}, diffusing probabilities within a 2-hop subgraph is enough to discover the important nodes around the given node. Besides, we set $k=4$ to guarantee that every node in the subgraph diffuses the probabilities at least twice to focus more probabilities on the nodes having denser structures.

Note that, in large networks, some nodes may have too many neighbor nodes, causing that conducting the diffusion operation on such nodes will dramatically increase the computational expense. To reduce the calculation of diffusion operation, we define a sampling operation for a given node that picks its 10 neighbor nodes with the highest clustering coefficient. Hence, instead of using all the nodes within the $l$-distance ego network for the given node, we apply a sampling operation on the given node to obtain at most 10 neighbor nodes, and then conduct sampling operation on each obtained neighbor node. Thus, the diffusion operation is limited to be performed inside the subgraph with at most 101 nodes (100 sampled nodes and a given node).

\subsection{HoSI Score of Subgraph to Node}
Analogically, the HoSI score of subgraph to node denotes the importance of a subgraph to a node. We introduce the definition of HoSI score of subgraph to node as follows.

\begin{Def}[HoSI Score of Subgraph to Node]
	Given a subgraph $G_{sub}=(V_{sub}, E_{sub})$, for any $u \in V$, the HoSI score of $G_{sub}$ to $u$ is calculated as follows: (1) apply diffusion operation on $u$ to obtain $p^{(k)}_{u,l}$; (2) sum up $HS(u,v)$ for all $v \in V_{sub}$, such that:
	\begin{equation}
		\label{eq_cn}
		HS(u,G_{sub}) = \sum_{v \in V_{sub}} HS(u,v),
	\end{equation}
	where $HS(u,G_{sub})$ denotes the HoSI score of $G_{sub}$ to $u$.
\end{Def}

For instance, as shown in Figure \ref{fig_2}, the HoSI scores of community (marked in pink) to node 1 and node 8 are 0.544 and 0.392, respectively. Consequently, node 1 is more likely to be the community member than node 8 in terms of the HoSI score.

\subsection{HoSI Score of Node}
The HoSI score of node is calculated by the total probabilities diffused from the neighborhood nodes to the given node. The definition of HoSI score of node is as follows.

\begin{Def}[HoSI Score of Node]
	For any $v \in V$, we apply the diffusion operation on each $u \in V_{v,l}$, where $V_{v,l}$ is the set of nodes within $G_{v,l}$, and the HoSI score of $v$ is given as: 
	\begin{equation}
		\label{eq_n}
		HS(v) = \sum_{u \in V_{v,l}} HS(u, v).
	\end{equation}
\end{Def}

For instance, in Figure \ref{fig_2}, the HoSI score of node 1 is calculated by $\sum_{i=2}^{7} HS(i, 1)$. Note that, due to the super nodes (i.e., the nodes having too many neighbor nodes) in large networks, the computational expense significantly increases. Thus, for any $v \in V$, we sample at most 100 nodes with the highest clustering coefficient from $V_{v,l}$ to calculate the HoSI score of $v$ that reduces the computational cost.

In this work, we use the HoSI score of node to identify the core members of communities because a node with high HoSI score denotes that there exists dense structures around the node. As a result, the nodes with higher HoSI score are more likely to be the core members of a local community than other nodes. Then, in this work, we define the core members as follows.

\begin{Def}[Core Members of a Community]
	The core members of a community are the few nodes with the highest HoSI scores inside the community.
\end{Def}

\section{HoSI Methodology}
In this section, we propose our method of HoSIM, which aims to detect multiple local communities containing the query node. There are three main stages in HoSIM, namely subgraph sampling, core member identification, and local community detection. The framework of HoSIM is shown in Algorithm \ref{alg_lorw}, and we elaborate each stage as follows. 

Note that, during the whole process, HoSIM may perform the diffusion operation on the same node many times. To deal with the issue, we store the probability vectors obtained by diffusion operation in a global data structure, which is shared in the whole process of HoSIM, to guarantee that the diffusion operation applies only once on the same node. Besides, we can also preserve the data structure in an external file. Then, the probability vectors can be repeatedly used when performing HoSIM to detect the local communities of other query nodes in the same network.

\subsection{Sampling a Subgraph}
In the first stage, HoSIM samples a subgraph for the query node that contains the most related nodes to the query node and the core members of the communities. 
There are two steps in the sampling process, and the framework is shown in Algorithm~\ref{alg_sam}. 

\begin{figure}[t]
	\centering
	\includegraphics[width=2.5in]{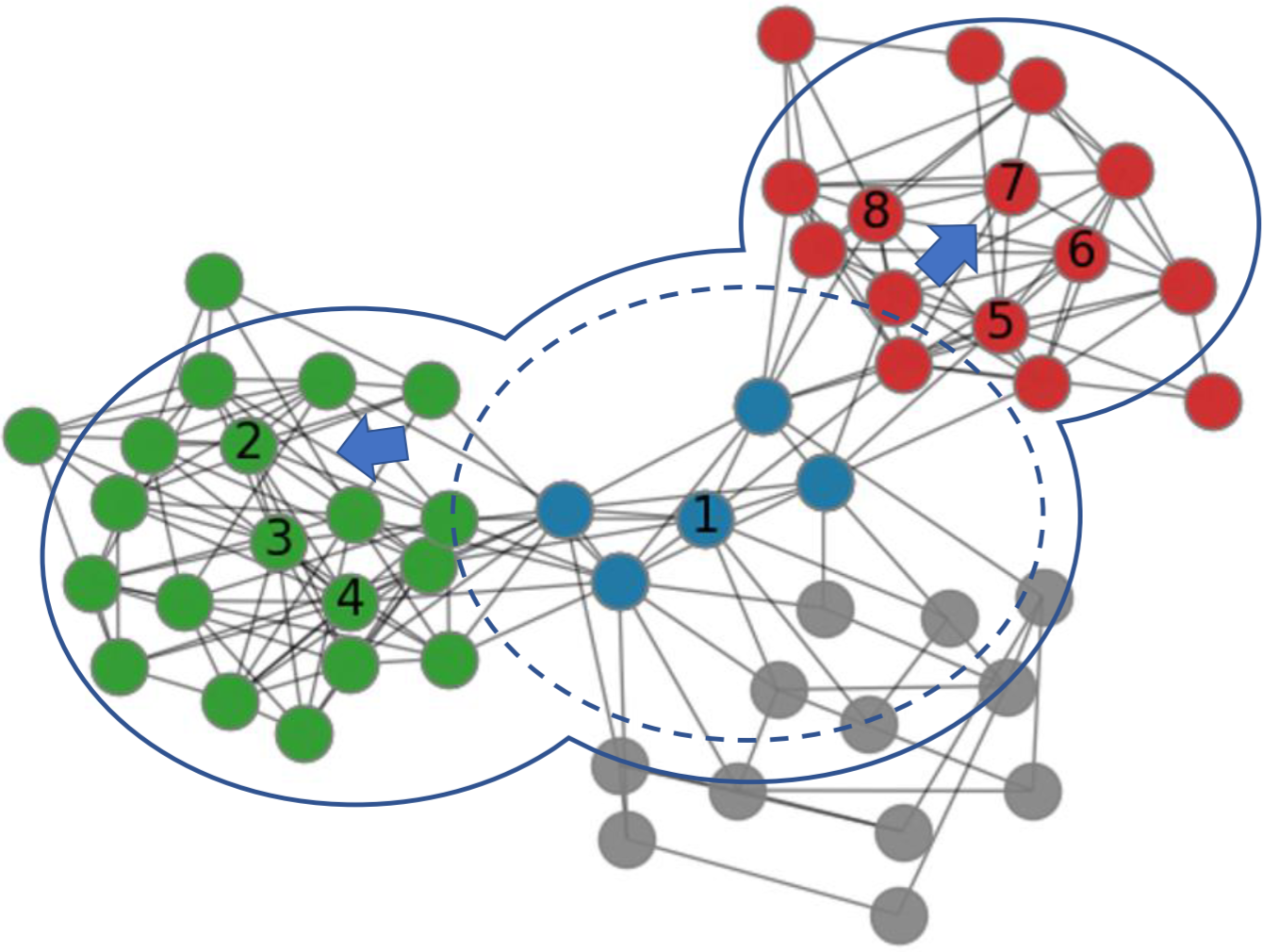}
	\vspace{-1em}
	\caption{Sampling strategy.}
	\label{fig_3}
\end{figure}

In the first step (lines 1 to 6), HoSIM iteratively performs breadth-first search (BFS) starting from the query node to obtain a subgraph, and the process stops when the size of subgraph is larger than a threshold $N_1$. Then, HoSIM applies ApproximatePR~\cite{AndersenCL06} to diffuse the probability from the query node within the subgraph. The top-$N_1$ nodes with the highest probability in the probability vector are selected to constitute the initial subgraph. For instance, as shown in Figure~\ref{fig_3}, the initial subgraph is within the blue dotted curve in which the nodes are most related to the query node 1.

In the second step (lines 7 to 17), HoSIM iteratively picks the neighbor nodes (i.e., the external adjacency nodes) of $G_{sub}$ and performs diffusion operation on each node in the neighbor nodes. HoSIM counts the HoSI score of $G_{sub}$ to each node in $neigh\_nodes$ according to Eq. (\ref{eq_cn}), respectively. Then, HoSIM sorts the nodes according to $hs\_sub\_node$ in descending order and adds top-$N_{iter}$ nodes into $G_{sub}$. This process finishes when the $add\_number$ is greater than a threshold $N_2$. For example, in Figure~\ref{fig_3}, the subgraph expands along the blue arrows, and the blue solid curve shows the final sampled subgraph. Note that, since the sizes of most ground-truth communities are less than 100, in this work, we set $N_1$ and $N_2$ both at 100, and $N_{iter}$ at 10. Therefore, the initial subgraph can be expanded for $\frac{N_2}{N_{iter}}=10$ rounds, and the size of the final subgraph is limited within $N_1 + N_2 = 200$.

Moreover, it is necessary to preserve the external adjacency nodes of the sampled subgraph because discovering a community relies on not only the members of the community but also the relations between the members and the external adjacency nodes (e.g., conductance). We denote the set of such external adjacency nodes as the shell of the subgraph, and HoSIM performs two-rounds of BFS from $G_{sub}$ to obtain the nodes as the shell of $G_{sub}$. Finally, HoSIM retains the subgraph for identifying the core members of communities and the shell of the subgraph for discovering the communities. 

\begin{algorithm}[t]
	\caption{HoSIM}
	\label{alg_lorw}
	\KwIn{graph $G$, query node $v_q$}
	\KwOut{$local\_communities$}
	$local\_communities = \emptyset$\;
	$G_{sub}, V_{shell}=sample\_subgraph(G, v_q)$\;
	$member\_sets=identify\_core\_members(G_{sub}, v_q)$\;
	\ForEach{$members \in member\_sets$}{
		$com=detect\_com(G_{sub}, V_{shell}, members, v_q)$\;
		add $com$ into $local\_communities$\;
	}
	\textbf{Return} $local\_communities$
\end{algorithm}

\begin{algorithm}[t]
	\caption{$sample\_subgraph$}
	\label{alg_sam}
	\KwIn{graph $G$, query node $v_q$}
	\KwOut{the sampled subgraph $G_{sub}$, and the shell of subgraph $V_{shell}$}
	$G_{sub}=[v_q]$\;
	\While{$|G_{sub}|<N_1$}{
		$G_{sub}=BFS(G_{sub})$\;
	}
	$prob\_vector = ApproximatePR(G_{sub},v_q)$\;
	$G_{sub}=pick\_top\_nodes(prob\_vector,N_1)$\;
	$add\_number = 0$\;
	\While{$add\_number<N_2$}{
		$neigh\_nodes=pick\_neighbor(G_{sub})$\;
		\ForEach{$node \in neigh\_nodes$}{
			$diffuse(node)$\;
		}
		$hs\_sub\_node=count\_hs\_1(G_{sub}, neigh\_nodes)$\;
		add top-$N_{iter}$ nodes from $hs\_sub\_node$ into $G_{sub}$\;
		$add\_number = add\_number + N_{iter}$\;
	}
	$V_{shell}=BFS(G_{sub}, 2)$\;
	\textbf{Return} $G_{sub}, V_{shell}$
\end{algorithm}

\subsection{Identifying the Core Members}
\begin{figure}[ht]
	\centering
	\includegraphics[width=2.5in]{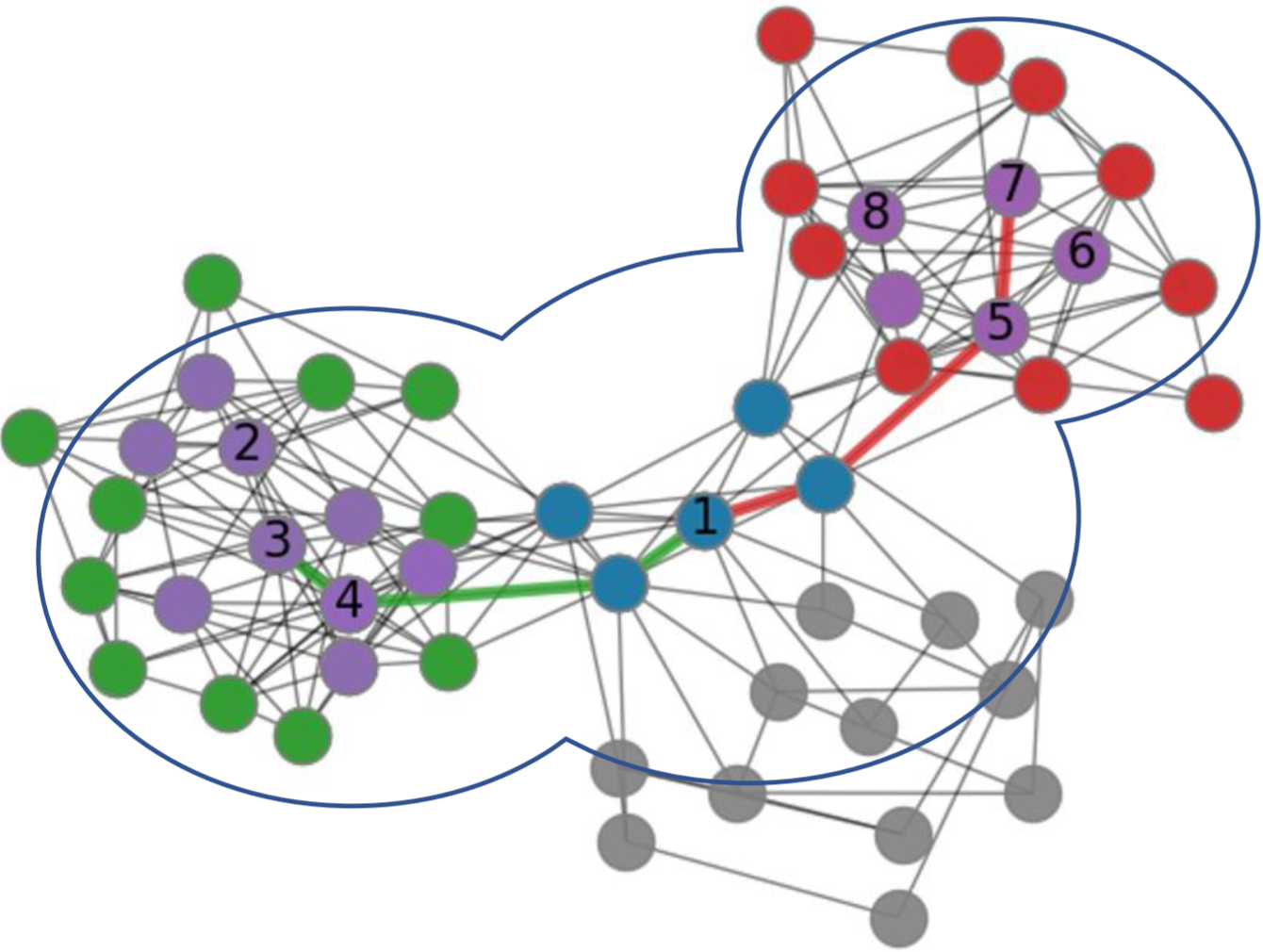}
	\caption{Identifying core members and picking seed nodes.}
	\label{fig_4}
\end{figure}

After sampling the subgraph, HoSIM identifies the core members of the communities within $G_{sub}$, and the process is shown in Algorithm~\ref{alg_find}. HoSIM applies the diffusion operation on each node in $G_{sub}$ and calculates the HoSI score of each node in $G_{sub}$ according to Eq. (\ref{eq_n}). Then, HoSIM selects the nodes whose HoSI score is greater than that of the query node as the core members. Finally, HoSIM discovers the independent connected components within $G_{sub}$ based on $core\_mems$. Meanwhile, the core members are divided into disjoint sets according to the independent connected components. As a result, HoSIM automatically determines the number of communities containing the query node. For instance, as shown in Figure~\ref{fig_4}, the purple nodes, which have greater HoSI score than the query node 1, are selected as the core members. Since the subgraph contains two independent connected components each including a set of purple nodes, we regard that query node 1 belongs to two communities.

Note that, for large and sparse networks, the core members of a community may not be connected, resulting that HoSIM may identify multiple sets of core members for the same community. Consequently, we calculate the sum of HoSI scores of all the nodes for each core member set and pick only 10 core member sets with the highest HoSI score. Then, the computational cost of detecting communities could be significantly reduced.

\begin{algorithm}[t]
	\caption{$identify\_core\_members$}
	\label{alg_find}
	\KwIn{sampled subgraph $G_{sub}$, query node $v_q$}
	\KwOut{the sets of core members $member\_sets$}
	\ForEach{$node \in G_{sub}$}{
		$diffuse(node)$\;
	}
	$hs\_node = count\_hs\_2(G_{sub})$\;
	$core\_mems = pick\_core\_mems(hs\_node, v_q)$\;
	$member\_sets = find\_comps(G_{sub}, core\_mems)$\;
	\textbf{Return} $member\_sets$
\end{algorithm}

\begin{algorithm}[t]
	\caption{$detect\_com$}
	\label{alg_det}
	\KwIn{sampled subgraph $G_{sub}$, shell of subgraph $V_{shell}$, core members $members$, query node $v_q$}
	\KwOut{the community $com$}
	$core\_node = pick\_core\_node(members)$\;
	$short\_path = explore\_path(G_{sub}, core\_node, v_q)$\;
	$seeds = pick\_seeds(short\_path, G_{sub}, core\_node)$\;
	$G_{union} = union(G_{sub}, V_{shell})$\;
	$com = PRN(G_{union}, seeds)$\;
	$com = add\_operation(com, G_{union}, \delta_{add})$\;
	$com = remove\_operation(com, \delta_{remove})$\;
	\textbf{Return} $com$
\end{algorithm}

\subsection{Detecting the Local Communities}
In the third stage, HoSIM discovers a community for each set of core members by three steps, and the framework is shown in Algorithm~\ref{alg_det}. 

First, to select a set of high quality seed nodes, HoSIM picks the node with highest HoSI score from the set of core members as core node and explores a shortest path starting from the query node to the core node within $G_{sub}$. Then, HoSIM considers the nodes along the shortest path and the neighbor nodes of core node within $G_{sub}$ as the set of seed nodes. In this way, HoSIM can accurately discover different communities based on the corresponding high quality seed set, respectively. For instance, in Figure~\ref{fig_4}, HoSIM explores a green path and a red path starting from the query node 1 to the core node 3 and core node 7, respectively, and the nodes along the path are contained in the corresponding set of seed nodes. 

In the second step, HoSIM performs PRN~\cite{AndersenCL06} with $\alpha=0.99$ and $\epsilon=0.001$ to grow the set of seed nodes into a community inside $G_{union}$, where $G_{union}$ is the subgraph induced by the nodes from $G_{sub}$ and $V_{shell}$. Additionally, to make fully use of the core node and the query node, we initialize higher probability weights on the core node and query node in PRN. Specifically, the probabilities are assigned to the core node, query node, and every other node as $\frac{0.2}{|seeds|} + 0.1$, $\frac{0.2}{|seeds|} + 0.7$, and $\frac{0.2}{|seeds|}$, respectively, where $|seeds|$ is the size of the initial seeds. Consequently, each of the core node and the query node has a higher probability to diffuse over $G_{union}$ than other seeds, resulting that the generated communities are more prone to cover the corresponding members, respectively.

Finally, HoSIM performs addition operation and removal operation in turn to further improve the quality of the community $com$. Specifically, addition operation and removal operation are defined as follows:
\begin{itemize}
	\item \textbf{Addition Operation} adds an external adjacency node of $com$ within $G_{union}$ into $com$ if the HoSI score of $com$ to the node is greater than the given threshold $\delta_{add}$. The process finishes when there does not exist any external adjacency node satisfying the condition.
	\item \textbf{Removal Operation} removes an internal node from $com$ if the HoSI score of $com$ to the node is less than the given threshold $\delta_{remove}$. The process finishes when there does not exist any internal node satisfying the condition.
\end{itemize}

\subsection{Complexity Analysis}
\label{sec_ana}
First, we calculate the complexity of diffusion operation. For each use of diffusion operation, the complexity is $O(|V_{u,l}|)$. In the worst case, HoSIM performs the diffusion operation on all nodes in the network, and the complexity is $O(|V|*|V_{u,l}|)$. By applying the sampling operation, the complexity is reduced to $O(|V|)$.

In the process of sampling subgraph, HoSIM applies ApproximatePR to sample an initial subgraph, and the complexity is $O(\frac{\log |V|}{\epsilon \alpha})$ in the worst case. In addition, according to Algorithm \ref{alg_sam}, the number of expansions is $N_2 / N_{iter}$, and for each expansion, the number of calculating the HoSI scores of a subgraph to a node is the size of the neighbors of $G_{sub}$. Thus, in the worst case, the complexity of calculating the HoSI scores of a subgraph to a node is $O(N_2 / N_{iter} * |V|)$. Since we fix the values of $N_2$ and $N_{iter}$, the complexity is $O(|V|)$.

In the process of detecting local communities, HoSIM conducts PRN to generate a local community within $G_{union}$ for each set of core members. In the worst case, the complexity of generating communities is $O(\frac{\log |V|}{\epsilon \alpha})$. Besides, for the addition operation and the removal operation in Algorithm~\ref{alg_det}, their complexities are both $O(|V|)$. Hence, the overall complexity of HoSIM is $O(|V| + \frac{\log |V|}{\epsilon \alpha})$.

\section{Experimental Results}
We conduct extensive experiments on synthetic networks and real-world networks to evaluate the effectiveness of HoSIM. Specifically, for each network, we pick a number of query nodes from all groups of nodes belonging to different number of communities, respectively. Then, we calculate the Jaccard $F_1$-score based on the ground-truth communities and the detected communities to evaluate the performances of algorithms. Besides, we also study the influence of parameters, the scalability of HoSIM, and the effectiveness of ARW.

The baselines include M-LOSP~\cite{HeSBHL15}, Multicom~\cite{HollocouBL17}, MLC~\cite{KamuhandaH18}, and SMLC~\cite{KamuhandaWH20}, and we use the default parameters for the baselines. All the experiments are performed on a processor: i5 @ 3.3 GHZ 3.3 GHZ, a RAM: 16 GB, and a 64-bit Windows operating system\footnote{Our codes are available at https://github.com/knyttstory/HoSIM.}.

\subsection{Comparison on Synthetic Networks}
\label{sec_lfr}
\begin{figure}[t]
	\centering
	\subfigure[$|V| = 10,000, \ \mu$ = 0.1]{
		\begin{minipage}[t]{0.5\linewidth}
			\centering
			\includegraphics[width=1.6in]{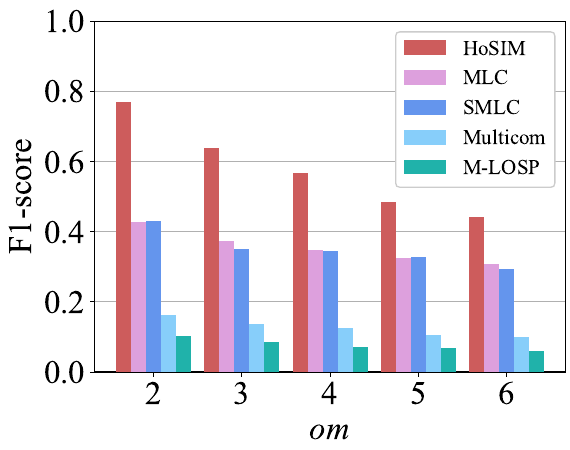}
		\end{minipage}
	}%
	\subfigure[$|V| = 10,000, \ \mu$ = 0.3]{
		\begin{minipage}[t]{0.5\linewidth}
			\centering
			\includegraphics[width=1.6in]{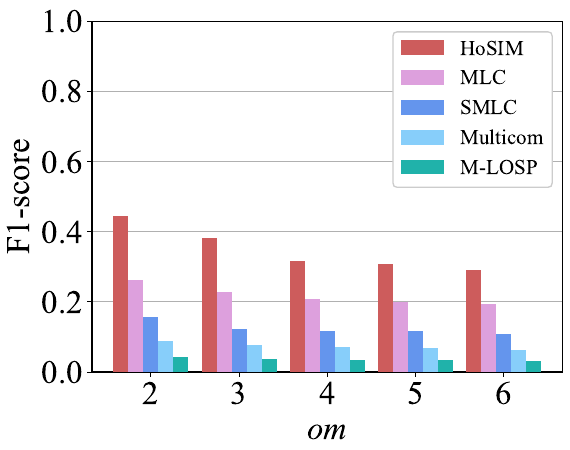}
		\end{minipage}
	}%
	
	\subfigure[$|V| = 100,000, \ \mu$ = 0.1]{
		\begin{minipage}[t]{0.5\linewidth}
			\centering
			\includegraphics[width=1.6in]{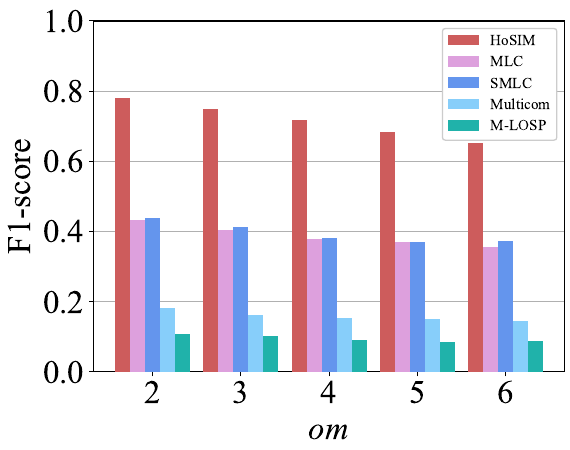}
		\end{minipage}
	}%
	\subfigure[$|V| = 100,000, \ \mu$ = 0.3]{
		\begin{minipage}[t]{0.5\linewidth}
			\centering
			\includegraphics[width=1.6in]{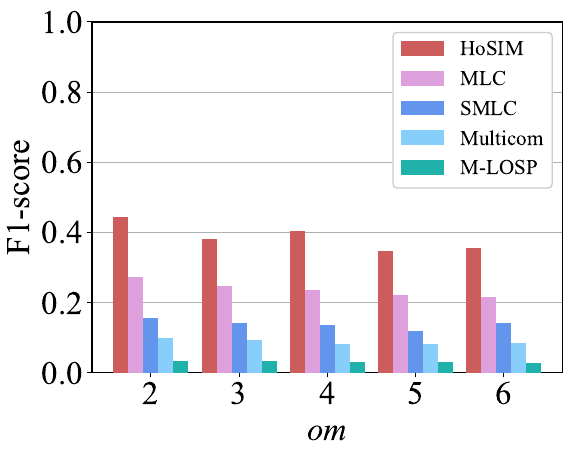}
		\end{minipage}
	}%
	\caption{Comparison of $F_1$-score on LFR networks.}
	\label{fig_LFR}
\end{figure}

LFR networks~\cite{lancichinetti2008benchmark} simulate the characteristics of real-world networks that have power-law distributions of node degrees and community sizes. We generate 20 unweighted and undirected LFR networks by controlling the number of nodes $|V|$, the mixing parameter $\mu$, and the memberships of overlapping nodes $om$. For each network, we pick 200 nodes that contains 100 nodes belonging to one community and 100 nodes belonging to $om$ communities. The detailed parameter setting is listed in Table \ref{tab_setl}. We set $\delta_{add}=0.3$ and $\delta_{remove}=0.2$ for HoSIM, and the results are shown in Figure~\ref{fig_LFR}.

\begin{table}[t]
	\small
	\centering
	\caption{Parameters for the LFR networks.}
	\label{tab_setl}
	\begin{tabular}{l|l}
		\hline\noalign{\smallskip}
		\makecell[c]{Parameter} & Description  \\
		\hline
		$|V| \in \{10,000; 100,000\}$ & Number of nodes \\
		$\mu \in \{0.1, 0.3\}$ & Mixing parameter \\
		$\bar{d}=10$ & Average degree \\
		$d_{max}=50$ & Maximum degree \\
		$|C| \in [20, 100]$ & Range of community size \\
		$\tau_1=2$ & Node degree distribution exponent \\
		$\tau_2=1$ & Community size distribution exponent \\
		$om \in \{2, 3, \cdots, 6\}$ & Overlapping membership \\
		$on=200$ & Number of overlapping nodes \\
		\noalign{\smallskip}\hline
	\end{tabular}
\end{table}

We can observe that HoSIM achieves all the highest $F_1$-scores, and the results of all the methods decay with the growth of $\mu$ or $om$. Since the core members of communities are more possible to be connected with the growth of $om$, HoSIM is difficult to precisely estimate the number of communities when $om$ is high. Consequently, the $F_1$-score of HoSIM decreases with the increase of $om$. On the other hand, the results of all the methods decay when $\mu$ increases to 0.3. It is because the structures of irrelevant nodes are more similar to that of the community members. Attributing to utilizing higher-order structural information, HoSIM remains the highest $F_1$-scores with much less declines than other methods.

\subsection{Comparison on Real-world Networks}
\label{sec_real}
We utilize Amazon network, DBLP network, and LiveJournal network, which are three datasets of real-world networks with ground-truth communities, downloaded from Stanford Network Analysis Project\footnote{http://snap.stanford.edu/}. For the three networks, we utilize the top 5,000 communities and remove their identical copies. The statistics of the networks are shown in Table \ref{tab_real}. Moreover, we randomly select 100 query nodes for each $om \in \{1, 2, \cdots, 5\}$ to accurately evaluate the performance of these methods. For HoSIM, 
we set $\delta_{add}=0.8$ and $\delta_{remove}=0.2$ on Amazon, and $\delta_{add}=0.7$ and $\delta_{remove}=0.1$ on DBLP, and $\delta_{add}=0.4$ and $\delta_{remove}=0.1$ on LiveJournal.

\begin{figure}[t]
	\centering
	\subfigure[Amazon]{
		\begin{minipage}[t]{0.5\linewidth}
			\centering
			\includegraphics[width=1.6in]{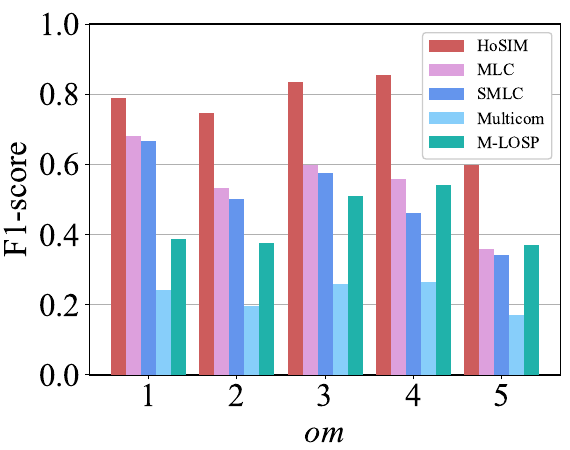}
		\end{minipage}
	}%
	\subfigure[DBLP]{
		\begin{minipage}[t]{0.5\linewidth}
			\centering
			\includegraphics[width=1.6in]{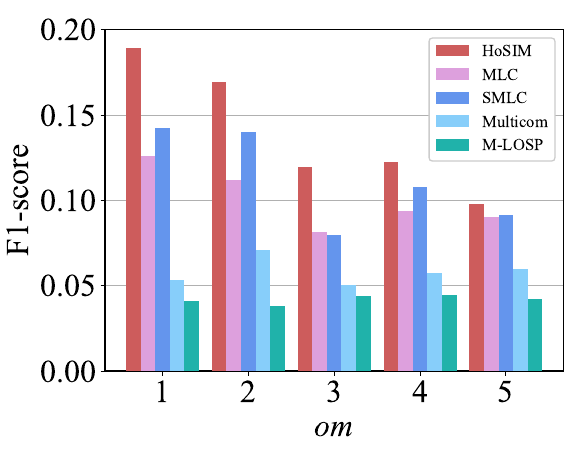}
		\end{minipage}
	}%
	
	\subfigure[LiveJournal]{
		\begin{minipage}[t]{0.5\linewidth}
			\centering
			\includegraphics[width=1.6in]{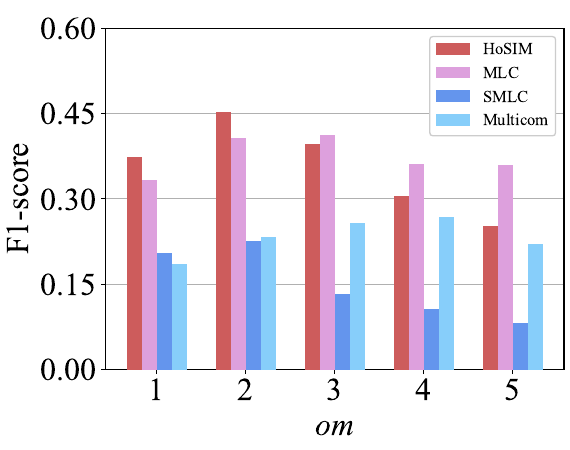}
		\end{minipage}
	}%
	\caption{Comparison of $F_1$-score on real-world networks.}
	\vspace{-1em}
	\label{fig_real}
\end{figure}

\begin{table}
	\small
	\centering
	\caption{Statistics of the real-world networks.}
	\label{tab_real}
	\begin{tabular}{l|cccc}
		\hline\noalign{\smallskip}
		Network & \#Nodes & \#Edges & \#Communities & $\mu$ \\
		\noalign{\smallskip}\hline\noalign{\smallskip}
		Amazon & \makecell[r]{334,863}& \makecell[r]{925,872}& 1,517 & 0.06 \\
		DBLP & \makecell[r]{317,080}& \makecell[r]{1,049,866}& 4,961 & 0.25 \\
		LiveJournal & \makecell[r]{3,997,962}& \makecell[r]{34,681,189}& 4,703 & 0.32 \\
		\noalign{\smallskip}\hline
	\end{tabular}
\end{table}

Figure \ref{fig_real} illustrates the results. Note that, we do not count the result of M-LOSP on LiveJournal network as it takes much longer time. We can observe that HoSIM achieves almost all the highest $F_1$-scores among the methods. Differing from LFR networks that the results steadily decline with the growth of $om$, the $F_1$-score fluctuates in real-world networks because their community structures are highly irregular. However, HoSIM is still able to effectively detect the communities by capturing the higher-order structural information. In addition, the results of all the methods on DBLP are much lower than that on Amazon and LiveJournal. It is because the value of $\mu$ on DBLP is high and some communities on DBLP are much larger and sparser. Thus, detecting a large community on DBLP is extremely difficult given only a single query node. 

\subsection{Parameter Study}
\begin{figure}[!t]
	\centering    
	\subfigure[$\mu=0.1$] {
		\label{fig_eff1}
		\includegraphics[width=2.0in]{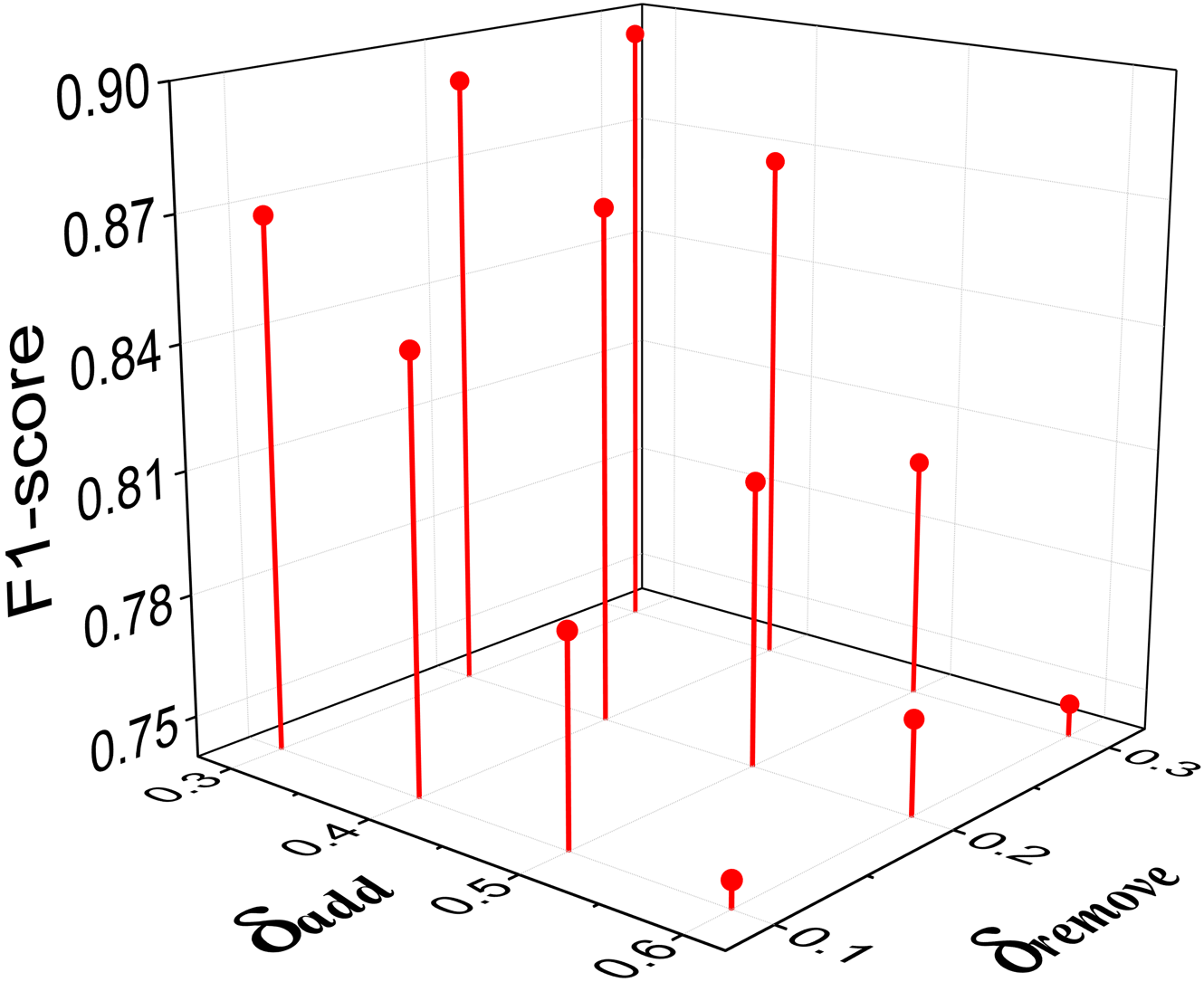}  
	}
	\subfigure[$\mu=0.3$] { 
		\label{fig_eff3}
		\includegraphics[width=2.0in]{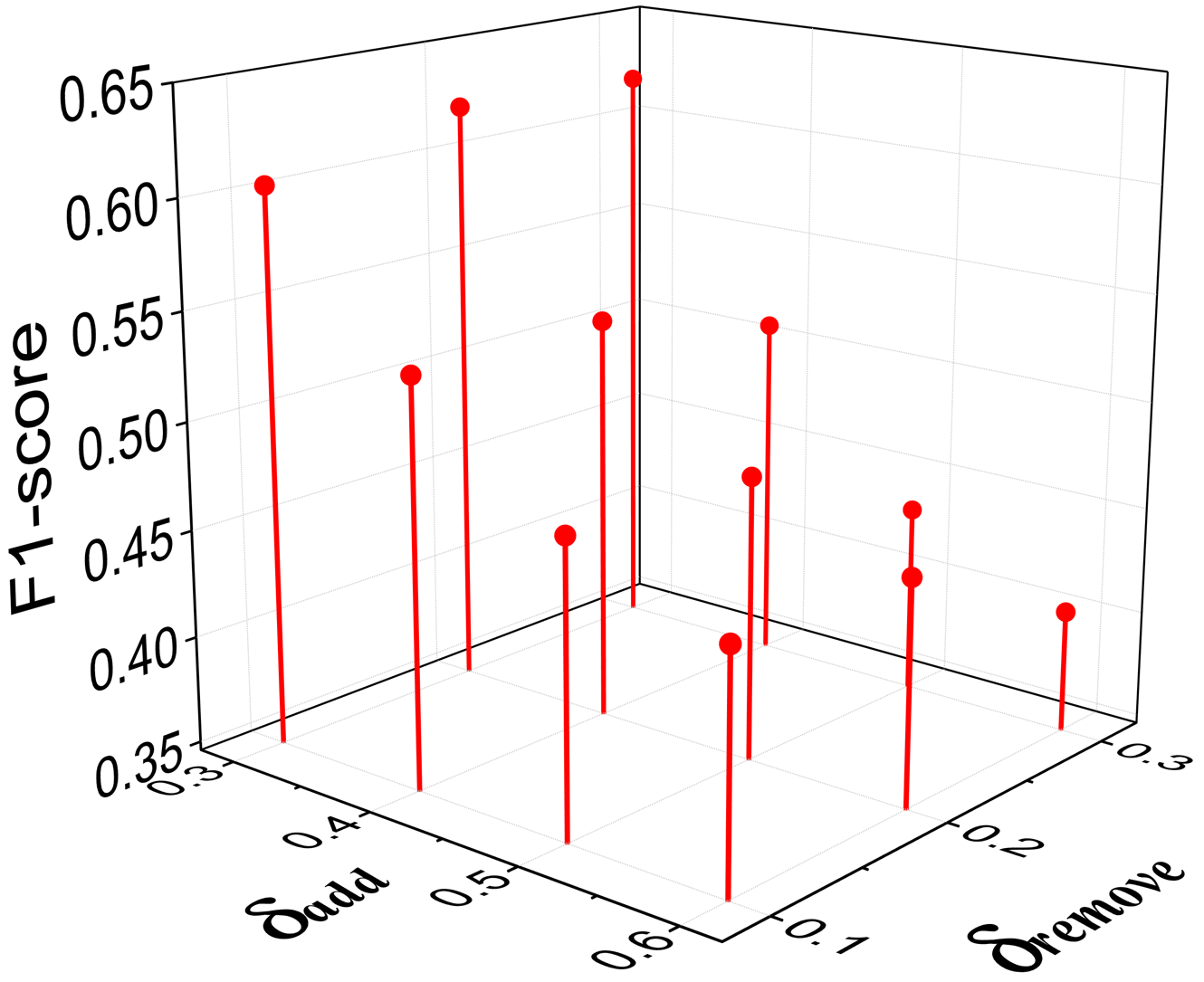}     
	}   
	\caption{Parameter study on $\delta_{add}$ and $\delta_{remove}$.}
	\label{fig_eff}     
\end{figure}

This subsection performs experiments to study the effects of the parameters, $\delta_{add}$ and $\delta_{remove}$. We utilize the LFR networks with $|V|=1,000$ and $om=2$. $\delta_{add}$ is set at 0.3, 0.4, 0.5, and 0.6, respectively, and $\delta_{remove}$ is set at 0.1, 0.2, and 0.3, respectively. The results are shown in Figure \ref{fig_eff}.

In Figure~\ref{fig_eff1}, we can find that the $F_1$-score always decays with the growth of $\delta_{add}$ because many community members cannot be joined into the communities when $\delta_{add}$ is high. Moreover, increasing the value of $\delta_{remove}$ steadily improves the detection results when $\delta_{add}$ is set at 0.3 or 0.4. It is because the irrelevant nodes are likely to be added into the communities when $\delta_{add}$ is low. Accordingly, increasing the value of $\delta_{remove}$ can effectively filter out irrelevant nodes from the communities. However, the communities do not contain enough members when $\delta_{add}$ is set at 0.5 or 0.6, resulting that the community members are possible to be excluded from the communities when $\delta_{remove}$ is 0.3.

In Figure~\ref{fig_eff3}, we can observe that the $F_1$-score declines with the growth of $\delta_{add}$ and $\delta_{remove}$ has little influence when $\delta_{remove}$ is low. It is because the structures around the irrelevant nodes are similar to that of the community members when $\mu=0.3$. Thus, $\delta_{remove}$ is insufficient to filter the communities to exclude irrelevant nodes. Furthermore, the detection results become obviously low when $\delta_{remove}$ is higher than 0.2 because several community members are removed from the communities. Consequently, we conclude that $\delta_{remove}$ should be cautiously increased when $\mu$ is high.

Note that, we also conduct experiments without applying addition operation and removal operation. The results are 0.6685 for $\mu=0.1$ and 0.4241 for $\mu=0.3$, respectively. Therefore, we can find that addition operation and removal operation can always improve the results when $\mu$ is small but may weaken the results when $\mu$ is big. The explicit study of setting $\delta_{add}$ and $\delta_{remove}$ on different networks will be our future work.

\subsection{Scalability}
This subsection illustrates the scalability of HoSIM based on the experiments on real-world networks. According to Section~\ref{sec_ana}, we count the average number of performing diffusion operation $N_{diff}$ and the average number of nodes used for each diffusion operation $N_{nodes}$. Besides, we compute the average number of nodes within $G_{sub}$ and the average number of nodes within $G_{union}$, denoted as $N_{sub}$ and $N_{union}$, respectively. The results are shown in Table~\ref{tab_sca}. In addition, we also count the average time of HoSIM and baselines, and the results are shown in Table~\ref{tab_scb}.

\begin{table}[ht]
	\small
	\centering
	\caption{Statistics of applying HoSIM on real-world networks.}
	\begin{tabular}{l|rrr}
		\hline
		& Amazon~ & DBLP~~ & LiveJournal~~ \\
		\hline
		$N_{diff}$ & 182.02~~ & 353.43~~ & 600.15~~ \\
		$N_{nodes}$ & 25.19~~ & 26.87~~ & 49.55~~ \\
		$N_{sub}$ & 172.23~~ & 182.75~~ & 116.19~~ \\
		$N_{union}$ & 968.01~~ & 5967.18~~ & 6762.30~~ \\
		\hline
	\end{tabular}
	\label{tab_sca}
\end{table}

\begin{table}[ht]
	\small
	\centering
	\caption{Average time (seconds) of applying algorithms on real-world networks.}
	\begin{tabular}{l|rrr}
		\hline
		\makecell[c]{Method} & Amazon~ & DBLP~~ & LiveJournal~~ \\
		\hline
		HoSIM & 0.15~~ & 0.30~~ & 1.80~~ \\
		MLC & 0.01~~ & 0.54~~ & 29.06~~ \\
		SMLC & 2.55~~ & 1.56~~ & 41.39~~ \\
		Multicom & 12.79~~ & 18.29~~ & 16.69~~ \\
		M-LOSP & 0.23~~ & 5.01~~ & -~~ \\
		\hline
	\end{tabular}
	\label{tab_scb}
\end{table}

In Table~\ref{tab_sca}, we can find that $N_{diff}$ and $N_{nodes}$ increase with the growth of the network scale. The values of $N_{union}$ on DBLP and LiveJournal are much bigger than Amazon because there are many super nodes (i.e., nodes with too many neighbor nodes) on DBLP and LiveJournal. However, $N_{sub}$ of LiveJournal is much smaller than Amazon and DBLP, denoting that HoSIM can effectively sample subgraphs for the query nodes on large networks. Additionally, Table~\ref{tab_scb} also shows that HoSIM has better scalability than other baselines.

\subsection{Effectiveness of ARW}
\begin{figure}[!t]
	\centering
	\subfigure[$\mu$ = 0.1]{
		\begin{minipage}[t]{0.5\linewidth}
			\centering
			\includegraphics[width=1.6in]{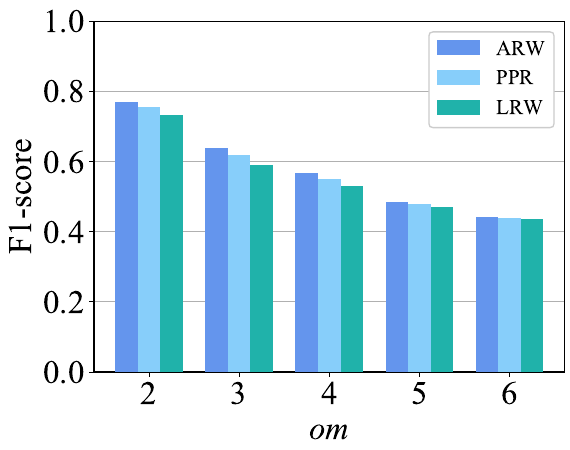}
			\label{fig_rw1}
		\end{minipage}
	}%
	\subfigure[$\mu$ = 0.3]{
		\begin{minipage}[t]{0.5\linewidth}
			\centering
			\includegraphics[width=1.6in]{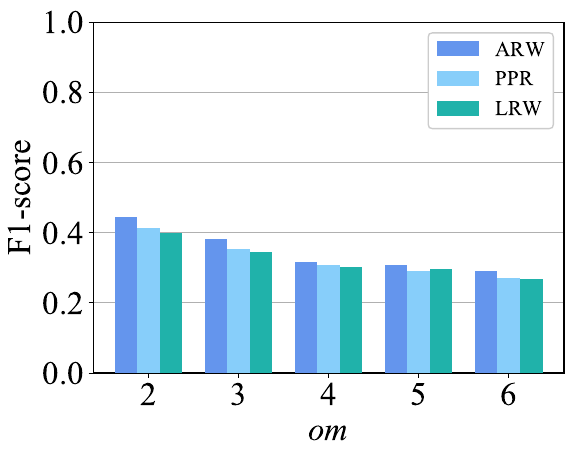}
			\label{fig_rw3}
		\end{minipage}
	}%
	\caption{Effectiveness of ARW.}
	\label{fig_rw}
\end{figure}

This subsection conducts experiments to study the effectiveness of ARW. We apply ARW, PPR, and LRW to implement the short random walks for the diffusion operation on the LFR networks with 10,000 nodes in Section~\ref{sec_lfr}, respectively. Figure \ref{fig_rw} illustrates the results. We can observe that ARW performs best in all cases, denoting that ARW is more effective to measure the HoSI score than PPR and LRW with short random walks. 

\section{Related Work}
\subsection{Single Local Community Detection}
Random-walk based models~\cite{AndersenCL06, KlosterG14, HeSBHL15, BianNCZ19} are widely applied in the problem of single local community detection. The rationale is to diffuse the probability from the seed node that measures the importance of neighborhood nodes to the seed node, and then pick a set of nodes with the minimum conductance as a local community. Some methods based on local modularity~\cite{LuoWP08, LuoZNL21} are proposed for detecting the local community. These approaches add an external node into the local community if the added node increases the local modularity of the local community. Zhang et al.~\cite{ZhangXYLWZY20} develop a semi-supervised algorithm called SEAL based on Generative Adversarial Networks (GANs) to discover the local community according to the features of community samples. In addition, there are many approaches proposed for finding the local community according to specific query requests~\cite{FangHQZZCL20}.

\subsection{Multiple Local Community Detection}
Random-walk based models can be flexibly extended to discover multiple local communities. M-LOSP~\cite{HeSBHL15} removes the query node from its ego network to obtain multiple independent connected components, and then applies LOSP~\cite{HeSBHL15} on each connected component to generate a local community. Multicom~\cite{HollocouBL17} utilizes random walk methods as scoring functions to generate the local embedding of the graph around the seed set, then iteratively selects new seeds and detects local communities. Kamuhanda et al.~\cite{KamuhandaH18, KamuhandaWH20} utilize Non-negative Matrix Factorization (NMF) to estimate the number of communities within the sampled subgraph and assign nodes into the communities. Ni et al.~\cite{NiLZH20} pick the seed nodes by using Nearest node with Greater Centrality (NGC) and fuzzy relation~\cite{LuoYBZ20}, then perform M~\cite{LuoWP08} and DMF\_F~\cite{LuoZJNH18} approaches to discover the local communities.

\subsection{Methods based on Higher-order Structure}
In recent years, there has been a growing interest in detecting communities based on the higher-order structural information~\cite{BensonGL16a, YinBLG17, TsourakakisPM17, LiHWL19, HuangCX20, SotiropoulosT21}. Jia et al.~\cite{JiaZ0W19} demonstrate that the occurrences of motifs are strongly associated with community structure. Consequently, making fully use of the high-order structural information greatly contributes to the discovery of communities. However, finding particular motifs in the network and estimating their influence are very difficult and computationally expensive. Hence, most existing community detection methods only consider the simplest triangular motif as the basis but ignore the effects of other important complex motif structures.

\section{Conclusion}
In this work, we address the problem of multiple local community detection based on HoSI. We first propose a new variant of random walk called ARW to effectively measure the HoSI score between nodes. Then, we present two new metrics to evaluate the HoSI score of a subgraph to a node and the HoSI score of node, respectively. By applying the first metric, we can calculate the importance of a community structure to a node in terms of higher-order structure, so that we can precisely judge whether a node belongs to the target community. By applying the second metric, we can determine whether there exists dense structures around a node and thus identify the core members of local communities.

Based on the proposed metrics, we present a new algorithm called HoSIM to detect multiple local communities for a given query node. The key idea is to utilize HoSI to find and identify the core members of communities and optimize the generated communities. Specifically, HoSIM effectively adds the core members of local communities into the sampled subgraph, identifies and divides the core members into disjoint sets, and improves the generated communities by adding and removing specific nodes.

\bibliographystyle{IEEEtran}
\bibliography{references}

\end{document}